\documentclass[epj,nopacs]{svjour}
\usepackage[utf8]{inputenc}
\usepackage[english]{babel}
\usepackage[numbers,compress]{natbib}
\usepackage{notoccite}
\usepackage{amssymb}
\usepackage{amsmath}
\usepackage{graphicx}
\usepackage{hyperref}
\usepackage{placeins}
\usepackage{xcolor}
\usepackage{soul}

\newcommand{\rmv}[1]{}
\newcommand{\rmvR}[1]{}
\newcommand{\com}[1]{}

\color{darkgray}
\begin{document}
\title{
	Influence of external magnetic field, finite-size effects and chemical potential on the phase transition of a complex scalar field
	}

\author{E. Cavalcanti\inst{1}\thanks{erich@cbpf.br}	\and
	E. Castro\inst{1}\thanks{erickc@cbpf.br} \and
	C.A. Linhares\inst{2}\thanks{linharescesar@gmail.com} \and
	A. P. C. Malbouisson\inst{1}\thanks{adolfo@cbpf.br}}
\institute{Centro Brasileiro de Pesquisas F\'{\i}sicas/MCTI,
	22290-180, Rio de Janeiro, RJ, Brazil 
	\and
	Instituto de F\'{\i}sica, Universidade do Estado do Rio
	de Janeiro, 20559-900, Rio de Janeiro, RJ, Brazil}

\abstract{A scalar model is built, as a quantum field theory defined on a toroidal topology, to describe a phase transition in films subjected to periodic boundary conditions and influenced by an external and constant magnetic field. Criticality is studied and the relations between the critical temperature, the film thickness, the magnetic field strength and the chemical potential are investigated. Since the model describes a second-order phase transition a comparison with the Ginzburg-Landau theory is made.
	\PACS{ {11.30.Qc}{} \and {11.10.Wx}{} \and {11.10.Kk}{}}
}

\maketitle

\section{Introduction}
Field theories defined on spaces with some of its dimensions compactified are interesting for several branches of theoretical physics. 
They can be related, for instance, to studies of finite-size scaling in phase transitions, to string theories or to phenomena involving extra dimensions in high and low-energy physics~\cite{cardy,polchinski,panilinha,pani1,pani3,pani6,pani6b,pani7,(g-2)NPB,claudio}. Finite-size effects are present in a wide range of thermal critical phenomena when the macroscopic dimensions of the system (volume, transversal section or length) are diminished, and the relevant thermodynamic quantities depend directly on the finite size of the system.

For a Euclidean $D$-dimensional space, compactification of $d$ coordinates with periodic boundary conditions means that its topology is of the type $\Gamma _{D}^{d}=(S^{1})^{d}\times \mathbb{R}^{D-d}$, with $1\leq d\leq D$. Each of these compactified dimensions has the topology of a circle $S^{1}$. We refer to $\Gamma _{D}^{d}$ as a toroidal topology. Mathematical foundations to deal with quantum field theories on toroidal topologies are consolidated in recent developments~\cite{PhysReport,AOP11,AOP09,TheBook}. This provides a general framework for results from earlier works as for instance in Refs.~\cite{NuclPhys2002,Abreu2006,Abreu2009,EPL2012,PRD2012,Emerson2013,PRD86,Linhares2006,Abreu2013}. This procedure is a generalization of the Matsubara imaginary-time formalism, used to introduce temperature in field theories. Basically, the inverse temperature $\beta$ is considered as another compactified coordinate. In the thermal field theory context, the Ginzburg-Landau action functional is important as a mathematical realization of the phase transition phenomenon, in which the field is interpreted as an order parameter linked to the particular microscopic physics of the system. The mass parameter governs the character of the transition, and so, finite-size and thermal effects are introduced together with corrections of the mass parameter. In this framework, we consider our system in the topology $\Gamma_D^d$ mentioned above, referring to a $D$-dimensional space with periodic boundary conditions on the imaginary time $\tau$ and on $d-1$ spatial coordinates. The Feynman rules are modified according to
\begin{subequations}
	\begin{align}
		\int \frac{dp_\tau}{2\pi} \rightarrow \frac{1}{\beta} \sum_{n=-\infty}^\infty, &\;p_\tau \rightarrow \frac{2\pi n}{\beta} - i\mu,\\
		\int \frac{dp_i}{2\pi} \rightarrow \frac{1}{L_i}\sum_{n=-\infty}^\infty , &\;p_i \rightarrow \frac{2\pi n}{L_i}, (i=1,\ldots,d-1),
	\end{align}
	\label{Eq:Prescricoes0}
\end{subequations}
\noindent where $\mu$ is the chemical potential. This is valid for the whole domain of temperatures, $0\leq \beta^{-1}<\infty$ and allows the study of finite-size effects.

In this article we consider a complex scalar model with self-interactive potential of the type $m^2\varphi^2+\lambda \varphi ^4$ ($\lambda >0$), which allows the system to undergo a second-order phase transition. In a pure field theoretical context, the action describes a system of complex self-interacting bosons in a external magnetic field. However, the Euclidean counterpart may represent two different systems: a charged Bose-Einstein condensate in the mean-field Gross-Pitaevskii functional format; or a superconductivity action functional. To build the effective action we use the non-perturbative technique of the loop expansion restricted, in our phenomenological approach, to an approximation of just one-loop corrections.
At finite temperatures, if we subject this system to a strong constant external magnetic field, the phenomenon of magnetic catalysis appears, as discussed in Refs.~\cite{Ayala,BoseEinstein01,BoseEinstein02}. It means that for strong magnetic fields the critical temperature increases with the growth of the external magnetic field.

Our main concern is to analyze two models within a field-theory approach, as used in statistical and con\-den\-sed-matter physics, and we apply the machinery of toroidal topologies to take into account finite-size effects. The thermal contribution is inserted through two different approa\-ches: (1) In a $D=3$ model, considering the Ginzburg-Landau approximation of a linear behavior of the mass term with the temperature, as employed for studies of superconducting films in a magnetic field background \cite{Linhares2006,PRB2002,Calan2004,Abreu2003}; (2) For $D=1+3$ in the framework of a quantum field theory with a toroidal topology \cite{PhysReport}, with two compactified dimensions ($d=2$), related to finite temperature and one compactified spatial coordinate, with compactification length $L$. From a condensed-matter point of view, we can think of this system as a heated film of thickness $L$, undergoing a phase transition under the influence of an applied magnetic field. A comparison between both models was also a subject of study in Ref.~\cite{Calza2016}.

The paper is organized as follows: At Section \ref{Sec:Model} we obtain the critical equation, given by the corrected mass term, to our models by use of the generalized formalism on toroidal topologies; we then explore the existence of a magnetically induced transition at zero-temperature. Section \ref{Results} is a collection of results derived from the analysis of the models, first studying the agreement between both ways to introduce temperature and then exploring further effects like existence of the minimal length, critical chemical potential and the zero-temperature phase diagram. In Section \ref{Sec:Conclusion} we present our conclusions. A quick demonstration that the model reproduces the zero magnetic field case is shown in Appendix \ref{Sec:LowOmega}.

\section{General model \label{Sec:Model}}
In this article we study systems that can be modeled by a complex scalar field $\varphi$. Its characteristic width, $L$, is introduced by a spatial compactification using the periodic boundary condition $\varphi (\tau;x,y,z) \equiv \varphi (\tau;x,y,z+L)$; the non-compactified coordinates are supposed to be very large so that it can be taken as extending to infinity. The system is considered to be in thermal equilibrium with a heat bath whose temperature is $T=\beta^{-1}$;temperature is introduced by the compactification $\varphi (\tau;x,y,z) \equiv \varphi (\tau+\beta^{-1};x,y,z)$ where $\tau$ is the imaginary time, as in Sec.~\ref{SecD4}, or by a direct linear dependence in the mass term, as in Sec.~\ref{SecD3}. 
In the general framework in a $D$-dimensional space, the proposed boundary conditions define a theory on a torus\cite{PhysReport} $\Gamma^2_D = (\mathbb{S}^1)^2 \times \mathbb{R}^{D-2}$, which is a suitable framework to treat thermal and finite-size effects. Here we consider the special case of Eqs.~\eqref{Eq:Prescricoes0} with just two compactifications ($d=2$).

The internal interaction is taken as a quartic self-interaction potential, for which we have a second-order phase transition. The Euclidean action of the system, expressed in a \textit{D}-dimensional space, is
	\begin{equation}
	S = \int d^D x \left\{ (D_\mu \varphi)^{\ast} D^\mu \varphi + m_0^2 \varphi^{\ast} \varphi + \frac{\lambda}{4}(\varphi^* \varphi)^2 \right\},
	\label{Eq:Action}
	\end{equation}
	\noindent where an external and constant magnetic field $B \hat z$ perpendicular to the system surface is introduced through minimal coupling, $\partial_\mu \rightarrow D_\mu = \partial_\mu - i e A_\mu$, and the gauge fixing is chosen so that $\mathbf{A} = B x \hat y$. After integration by parts we get for the free action an expression of the type $-\int d^{D}r\,\varphi ^{\ast}	\mathcal{D}\varphi $, where the differential operator becomes
	\begin{equation}
	\mathcal{D} = \nabla^2 - m_0^2 - 2i\omega x \partial_y -\omega^2 x^2, \label{operator}
	\end{equation}
	\noindent and $\omega=eB$ is the so-called cyclotron frequency. Note that the free action carries the external magnetic field. The free propagator $G(\mathbf{r}, \mathbf{r'})$ can then be written in terms of the eigenvalues and eigenfunctions of the operator $\mathcal{D}$ (see Ref.~\cite{Lawrie}). 
	
	In this article, since we employ the lowest approximation of the loop expansion, it suffices to have the knowledge of $G(\mathbf{r}, \mathbf{r})$, which can be obtained by taking the coincidence limit, $\mathbf{r}=\mathbf{r}^{\prime}$,
	\begin{equation}
	G(\mathbf{r},\mathbf{r}) = \int \frac{d^{D-2} q}{(2\pi)^{D-2}} \frac{\omega}{2\pi} \sum_{\ell=0}^{\infty} \frac{1}{\mathbf{q}^2 + (2\ell+1)\omega + m^2_0},
	\label{Eq:Propagator}
	\end{equation}
	\noindent where the sum over $\ell$ indicates the contribution of all Landau levels.	Note that the free propagator has a manifest dependence on the magnetic field, and therefore also the 1-loop corrections.
	As a consequence, to introduce temperature and finite-size effects in the critical equation we are restricted to perform compactifications on the remaining $D-2$ coordinates. We consider two cases, the first one with $D=1+3$, that has thermal effects introduced using imaginary time; the second one with $D=3$ where the thermal effects are introduced directly into the mass term.
	
	The phase transition can be identified from a change in the value of the order parameter (vacuum expectation value of the field $\varphi$); in the disordered/symmetric phase $\varphi(\mathbf{r})=0$, in the ordered/broken phase $\varphi(\mathbf{r})\neq0$. The phase transition is related to a change in the sign of the concavity of the effective action at the minimum $\phi(\mathbf{r})=0$. The concavity is given by the second derivative of the effective action with respect to the field,
		\begin{equation*}
		C^{(\varphi^\star)}=	\left.\frac{\delta^2\Gamma_{\text{eff}}(\varphi)}{\delta \varphi(\mathbf{r_1})\delta \varphi^\star(\mathbf{r}_2)}\right|_{\varphi=\varphi^\star}.
		\end{equation*}
	The effective action can be expanded in loops. At the 1-loop approximation it is given by the set of all one-loop diagrams with an arbitrary number of $\varphi^2$ insertions. At $\varphi(\mathbf{r})=0$ the only contribution that survives corresponds to a single $\varphi^2$ insertion and the concavity, $C^{(0)}$, is
	\begin{multline}
	\left.\frac{\delta^2\Gamma_{\text{eff}}(\varphi)}{\delta \varphi(\mathbf{r_1})\delta \varphi^\star(\mathbf{r}_2)}\right|_{\varphi=0} = 
	(-\nabla^2 + 2i\omega x \partial_y + \omega^2 x^2+m_0^2)\\\times \delta(\mathbf{r}_1 - \mathbf{r}_2) + \frac{\lambda}{2} G(\mathbf{r}_1, \mathbf{r}_1) \delta(\mathbf{r}_1 - \mathbf{r}_2). \label{Eq:DerivadaFuncional}
	\end{multline}
	We take the first three terms inside parentheses as kinetic terms. In analogy with systems without external magnetic field, the effective mass is defined through the removal of these kinetic terms and the Dirac delta function,
	\begin{equation}
	m_{\text{eff}}^2 = m^2_0 + \frac{\lambda}{2} G(\mathbf{r}, \mathbf{r}).
	\end{equation}
	\noindent In this way, it follows that the concavity analysis -- and, by consequence, the phase transition study -- is in correspondence with the analysis of the sign of $m_{\text{eff}}^2$. 
	
	To treat the propagator~\cite{PhysReport} we apply the prescriptions of Eq.~\eqref{Eq:Prescricoes0} into Eq.~\eqref{Eq:Propagator}, then solve the remaining momentum integrations by use of dimensional regularization\cite{DimReg,DimReg2} and finally relate the infinite sums to the Epstein-Hurwitz spectral zeta functions. These have representations in the whole complex plane in terms of the modified Bessel functions of the second kind $K_\nu$ (see Ref.~\cite{ElizaldeRomeo,elizaldebook}), 
\begin{multline}
m_{\text{eff}}^2 = m_0^2 
+ \frac{\lambda \omega\Gamma\left(\frac{4-D}{2}\right)}{(4\pi)^\frac{D}{2} (2\omega)^{\frac{4-D}{2}}} 
\zeta_H\left(\frac{4-D}{2},\frac{m_0^2}{2\omega}+\frac{1}{2}\right)\\
+ \frac{\lambda}{(2\pi)^{\frac{D}{2}}} \sum_{\ell=0}^{\infty} \frac{\omega}{m_\ell^{\frac{4-D}{2}}} \Bigg\{
\sum_{n=1}^{\infty} \frac{\cosh(n \beta \mu)}{(n \beta)^{\frac{D-4}{2}}} K_{\frac{4-D}{2}} \left( n \beta m_\ell \right)\\
+ \sum_{n=1}^{\infty} \frac{K_{\frac{4-D}{2}} \left( n L m_\ell \right)}{(n L)^{\frac{D-4}{2}}} 
+ 2 \sum_{n_0,n_1=1}^{\infty} \frac{\cosh(n_0 \beta \mu)}{(n_0^2 \beta^2 + n_1^2 L^2)^{\frac{D-4}{4}}} \times \\ \times K_{\frac{4-D}{2}} \left(m_\ell \sqrt{n_0^2 \beta^2 + n_1^2 L^2} \right)
\Bigg\}.
\label{Eq:MainM2}
\end{multline}	
In the last equation we have defined $m_\ell^2 = m_0^2 + (2\ell+1)\omega$ and used the Hurwitz zeta function $\zeta_H(s,a) = \sum_{\ell=0}^{\infty}\frac{1}{(\ell+a)^s}$.

We choose, as a way to compare two different models ($D=1+3$ and $D=3$), to work with dimensionless parameters. Defining some arbitrary mass scale $\xi$ we have
\begin{align*}
\varphi = \xi ^{\frac{D-2}{2}} \widetilde{\varphi},\quad
m = \xi \widetilde{m},\quad
\lambda = \xi^{4-D} \widetilde{\lambda},\quad
\omega = \xi^2 \widetilde{\omega},\\
T = \xi^{-1} \widetilde{T},\quad
L = \xi^{-1}\widetilde{L},\quad
\mu = \xi \widetilde{\mu};
\end{align*}
\noindent We then expect that the dimensionless parameters can be used to compare the models. As the scale $\xi$ only contributes as a global term, the final expressions do not depend on it. In what follows we drop the tilde and all parameters are understood be dimensionless unless explicitly stated. 

\subsection{$D=1+3$ \label{SecD4}}

In this section we explore the case $D=1+3$, constructing its critical equation.  As a consequence of the dimension there is a pole in the gamma function, see Eq.~\eqref{Eq:MainM2}, which can be made explicit by taking $D=4-2\varepsilon$ for vanishing $\varepsilon$, in such a way that
\begin{multline*}
\frac{\lambda \omega\Gamma\left(\frac{4-D}{2}\right)\zeta_H\left(\frac{4-D}{2},\frac{m_0^2}{2\omega}+\frac{1}{2}\right)}{(4\pi)^\frac{D}{2} (2\omega)^{\frac{4-D}{2}}} 
\stackrel{D=4-2\varepsilon}{=} \frac{\lambda\omega}{(4\pi)^2} \Bigg\{
\Bigg[\frac{1}{\varepsilon} -\gamma \\+ \ln\left(\frac{2\pi}{\omega}\right)\Bigg] \zeta_H \left(0, \frac{m_0^2+\omega}{2\omega}\right)
+ \zeta_H^{(1,0)} \left(0, \frac{m_0^2+\omega}{2\omega}\right)
\Bigg\} =\\
 \frac{\lambda\omega}{(4\pi)^2} \left\{
-\frac{m_0^2}{2\omega} \ln \left(\frac{2\pi}{\omega}\right)
+ \ln\frac{\Gamma\left(\frac{m_0^2}{2\omega}+\frac{1}{2}\right)}{\sqrt{2\pi}}\right\}.
\end{multline*}
\noindent In the last line of the above equation the divergences are removed employing the $\overline{MS}$ scheme and we use the representations of the Hurwitz zeta and its derivative ($\zeta_H^{(1,0)} (a,z) = \frac{\partial}{\partial a} \zeta_H (a,z)$) at $a=0$. We point out that the expression is valid for $m_0^2 +\omega>0$. Finally we obtain from Eq.\eqref{Eq:MainM2} that
\begin{multline}
m_{\text{eff}}^2 = m_0^2 
+ \frac{\lambda\omega}{(4\pi)^2} \left\{
-\frac{m_0^2}{2\omega} \ln \left(\frac{2\pi}{\omega}\right)
+ \ln\frac{\Gamma\left(\frac{m_0^2}{2\omega}+\frac{1}{2}\right)}{\sqrt{2\pi}}\right\} \\
+ \frac{\lambda \omega}{(2\pi)^{2}} \sum_{\ell=0,n=1}^{\infty} \Bigg[
\cosh(n \beta \mu) K_{0} \left( n \beta m_\ell \right)
+ K_{0} \left( n L m_\ell \right) \Bigg]\\
+ \frac{2\lambda \omega}{(2\pi)^{2}} \sum_{\ell=0}^{\infty} \sum_{n_0,n_1=1}^{\infty} \cosh(n_0 \beta \mu) K_{0} \left( \sqrt{n_0^2 \beta^2 + n_1^2 L^2}m_\ell \right).
\label{Eq:MassD4}
\end{multline}

\noindent In the limit $\omega\rightarrow 0$ we recover the theory in the absence of external field. As a example, in the bulk (see Appendix \ref{Sec:LowOmega}), the first order correction in $\omega$ is
\begin{multline}
m_{\text{eff}}^2 = m_0^2 
+\frac{\lambda m_0^2}{2(4\pi)^2} \left[\ln \frac{m_0^2}{4\pi} -1\right] -\frac{\lambda  \omega ^2}{12(4\pi) ^2 m^2} \\
+ \frac{\lambda}{(2\pi)^{2}}\Bigg\{
\sum_{n=1}^{\infty} \frac{m_0 T}{n} \cosh(n \beta \mu) K_{1}\left(\frac{m_0 n}{T}\right) \\- \frac{\omega^2 n m_0}{24 T} K_1\left(\frac{m_0 n}{T}\right)\Bigg\}
+ \mathcal{O}(\omega^4).
\label{Eq:MassD4omega0}
\end{multline}
\noindent Although analytically we can make $\omega\rightarrow0$ this procedure is not numerically simple as the sum over the Landau levels converges slowly for an almost null magnetic field. This statement is more evident in Fig.~\ref{fig:qbulk}, where the point represented by the full black diamond is the $\omega=0$ limit, while the dotted curve is the one obtained for generic $\omega$. We emphasize that, although the convergence is faster for high values of the magnetic field, our model is not restricted to that situation.

\subsection{$D=3$ \label{SecD3}}

To construct the $D=3$ model we take Eq.~\eqref{Eq:MainM2} for $d=1$ (removing $\beta$-dependent terms), specify $D=3$ -- a case where there is no pole -- and then sum over $n$ (size-dependent frequencies) to obtain
\begin{multline}
m_{\text{eff}}^2 = m_0^2 
+ \frac{\lambda \sqrt{\omega}}{8\sqrt{2}\pi} \zeta_H\left(\frac{1}{2},\frac{m_0^2+\omega}{2\omega}\right)
\\+ \frac{\lambda \sqrt{\omega}}{4\sqrt{2}\pi} \sum_{\ell=0}^{\infty} \frac{\left[e^{ \sqrt{2\omega}L \sqrt{\ell + \frac{m_0^2+\omega}{2\omega}}}-1\right]^{-1}}{\sqrt{\ell + \frac{m_0^2+\omega}{2\omega}}}.
\label{Eq:MassD3}
\end{multline}

We see that the corrected mass $m_{\text{eff}}^2$ now depends on $L$. The temperature is introduced via the mass term as $m_0^2 = \alpha(T-T_0)$ and $m_{\text{eff}}^2 = \alpha(T-T_c)$, corresponding with the usual choice of the Ginzburg-Landau theory of superconductivity. $T_c$ is the critical temperature and $T_0$ is the critical temperature in the bulk ($L\rightarrow\infty$) with no magnetic field ($\omega=0$).

In the bulk , the third term disappears and we can determine an analytic limit for the condition $\omega\to 0$. In this limit the Hurwitz zeta function has the following asymptotic behavior
	\begin{equation}
	\zeta_{H}\left(\frac{1}{2},\frac{\omega+m^{2}}{2\omega}\right)\sim-2\left(\frac{\omega+m^{2}}{2\omega}\right)^{\frac{1}{2}}\stackrel{\omega \to 0}{=}-\sqrt{\frac{2}{\omega}}|m|,
	\end{equation}
\noindent therefore,
\begin{equation}
	m_{\text{eff}}^{2}=m_0^{2}-\frac{\lambda}{8\pi}|m_0|,
\end{equation}
\noindent which fixes a first correction to the critical temperature in this regime,
	\begin{equation*}
	T_c = T_0 + \frac{\lambda^2}{64\pi^2 \alpha}.
	\end{equation*}
	
\subsection{Magnetically induced non-thermal transition}

In this section we consider the zero-temperature limit and explore the existence of the so-called quantum phase transition\cite{Mucio2007,Muciobook}, which are phase transitions driven not by the temperature but by another parameter.In our case, as the effective mass parameter (Eq.~\eqref{Eq:MassD4} or Eq~.\eqref{Eq:MassD3}) depends on the magnetic field, there can be a magnetically induced phase transition. In other words, a first question to be asked is: Is there any magnetically induced quantum phase transition undergoing in the system? In what follows, we shall make this investigation in the bulk ($L\rightarrow\infty$) with no chemical potential ($\mu=0$). 

\begin{figure}[t]
	\centering
	\includegraphics[width=0.7\linewidth]{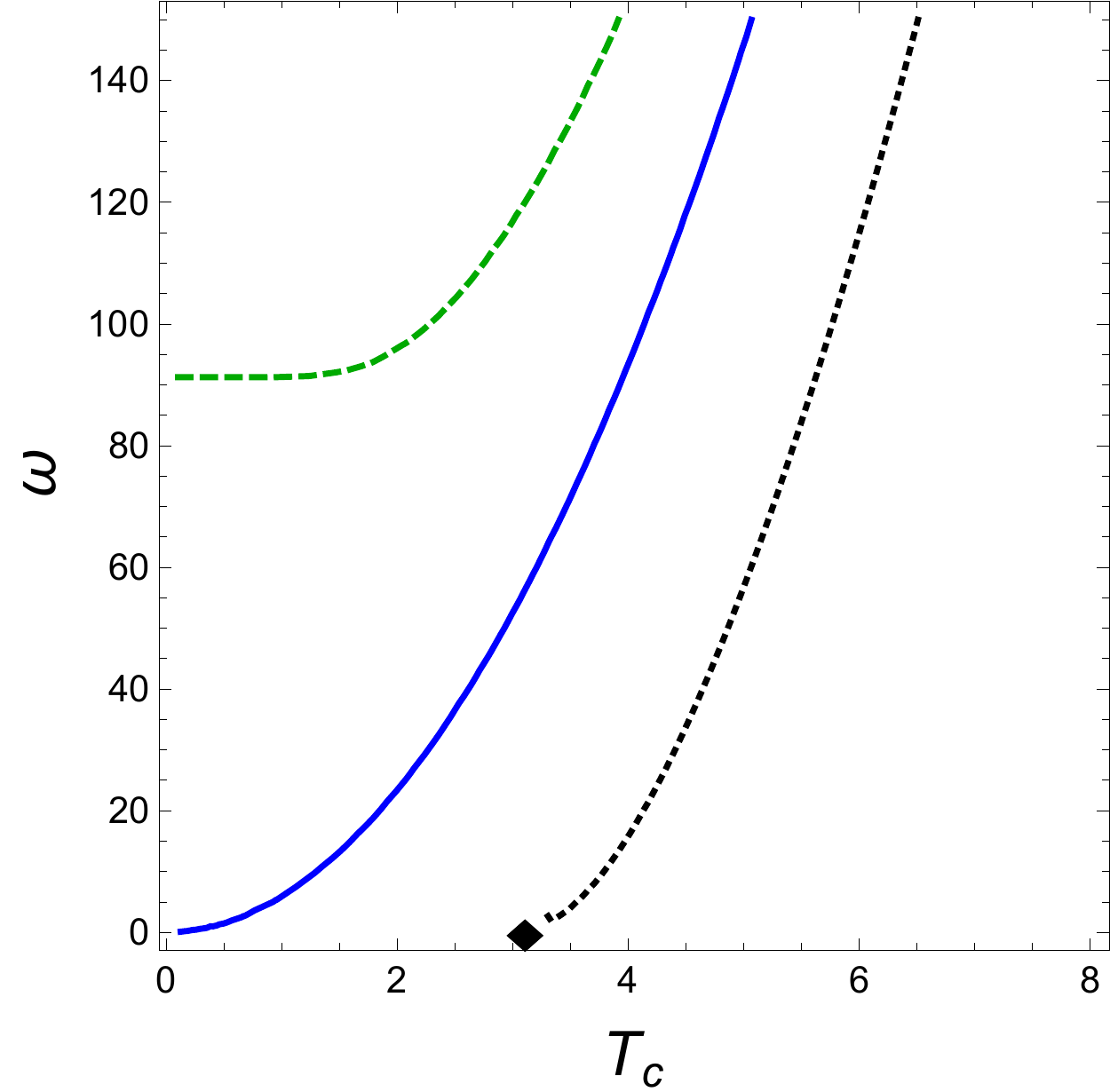}
	\includegraphics[width=0.7\linewidth]{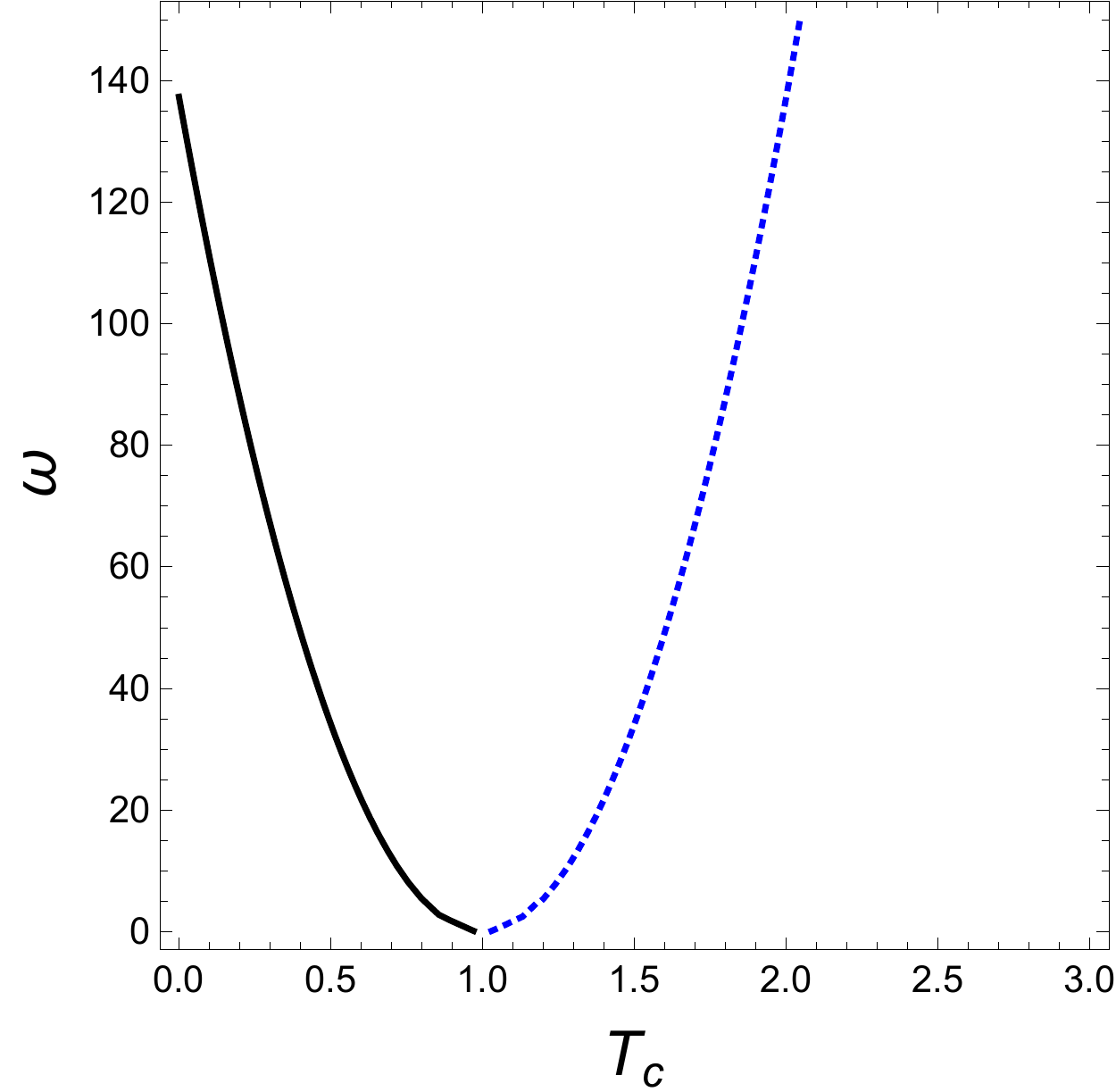}
	\caption{Phase diagrams at the bulk for $D=1+3$ (top) and $D=3$ (bottom). For $D=1+3$ we have $\lambda=0.5$ and squared mass $m_0^2=-0.2, 0, 0.1$ respectively for the dotted, full and dashed curves. The ordered phase is above the curves. The black diamond is the critical temperature for $\omega=0$, $m^2=-0.2$. For $D=3$, $\lambda=1$ and $T_0=1$. The black full curve has $\alpha=-0.2$, with the ordered phase above the it. The dotted blue curve has $\alpha=0.2$ with the ordered phase above it.}
		\label{fig:qbulk}
\end{figure}

For the $D=1+3$ model, Eq.~\eqref{Eq:MassD4}, the mass term at zero-temperature and in the bulk regime is just
\begin{equation*}
m_{\text{eff}}^2 = m_0^2 
+ \frac{\lambda\omega}{(4\pi)^2} \left\{
-\frac{m_0^2}{2\omega} \ln\left(\frac{2\pi}{\omega}\right) 
+ \ln\frac{\Gamma\left(\frac{m_0^2}{2\omega}+\frac{1}{2}\right)}{\sqrt{2\pi}} \right\}.
\end{equation*}
\noindent As an approximation, we can take the asymptotic behavior with respect to the magnetic field using
\begin{align*}
\lim\limits_{\omega\rightarrow\infty} \frac{\ln 1/\omega}{\omega} &= 0,\\
\lim\limits_{\omega\rightarrow\infty} \ln\Gamma\left(\frac{a}{\omega}+\frac{1}{2}\right)&= \frac{\ln \pi}{2},
\end{align*}
\noindent in such a way that

\begin{equation}
m_{\text{eff}}^2 = m_0^2 - \frac{\lambda \omega \ln 2}{2 (4\pi)^2}.
\end{equation}

This expression, although exact only for high values of $\omega$, is in a very good agreement with the exact expression. In means that there exists a critical field $\omega_c$ (which defines a null critical temperature, $m_{\text{eff}}^2(T_c=0,\omega_c)=0$) at the bulk \textit{only} if $m_0^2>0$. This behavior should already be expected, taking into account the bulk case where only the external magnetic field $\omega$ and the temperature $T$ play a role: while $\omega$ \textit{induces} order in the system (for $m_0^2>0$ and fixed $T$ there always exists a $\omega_{c}$ such that $m_{\text{eff}}^{2}<0$ for $\omega>\omega_{c}$), $T$ \textit{induces} disorder (for a fixed magnetic field, there always exists a $T_{c}$ such that  $m_{\text{eff}}^{2}>0$ for $T>T_{c}$). If the system is already ordered at zero-temperature ($m_0^2<0$) there can only be thermal transitions. But, if the system is disordered at zero-temperature ($m_0^2>0$) the increase in magnetic field can generate ordering, so the system undergoes a non-thermal transition. 

However, the temperature and the magnetic field begin to compete above $\omega_c$, meaning that it is always possible to obtain some critical temperature at which disorder is restored. We show the general behavior in Fig.~\ref{fig:qbulk} for a non-zero critical temperature, the \textit{competing} behavior is evident and signals the so-called magnetic catalysis: as the external magnetic field grows the critical temperature increases. For $m_0^2 = -0.2$ the sum over Landau levels is slowly convergent for low-valued magnetic fields. However, the zero field behavior is well known, Eq.~\eqref{Eq:MassD4omega0}, and is exhibited as a black diamond (see Fig.~\ref{fig:qbulk}) which agrees with the curve behavior.

For $D=3$ the magnetically induced non-thermal phase transition happens in a different context. Taking the model of Eq.~\eqref{Eq:MassD3}, in the bulk and with zero-temperature and then expanding it asymptotically for high values of $\omega$, we obtain
\begin{multline*}
M^2 = - \alpha T_0 + \frac{\lambda \sqrt{\omega}}{8\sqrt{2}\pi} \Big[(\sqrt{2}-1)\zeta(1/2)\\+(\sqrt{2}-1/2)\zeta(3/2)\frac{\alpha T_0}{2\omega} + \mathcal{O}(\omega^{-2})\Big].
\end{multline*}
Where $\zeta(a)$ is the Riemann zeta function. Critically, this means, considering only the first correction,
\begin{equation}
\sqrt{\omega_c} = \frac{8 \sqrt{2}\pi}{(\sqrt{2}-1)\zeta(1/2)} \frac{\alpha T_0}{\lambda},
\end{equation}
\noindent and as $\zeta(1/2)<0$, the condition $\alpha<0$ must be fulfilled to guarantee the existence of a magnetically induced transition.

We stress that, similarly to the $D=1+3$ case, the existence of a magnetic transition at zero-temperature requires that the phase is originally disordered (as the increase in magnetic field stimulates the ordering). Here, as $m_0^2 (T=0) = -\alpha T_0$ this can only happens if $\alpha<0$. 

The $\alpha<0$ case is strange as it describes an inverse transition: the increase in temperature orders the system. We exhibit this result as an academic curiosity, see Fig.~\ref{fig:qbulk}, for both possible situations: $\alpha<0$, $\alpha>0$. The common point to both curves is $T_0$, which is the critical temperature value at $\omega=0$.

\section{Comparison and results \label{Results}}

In this section we explore the consequences of the discussed models. As a general guide we assume that the phase is ordered/broken at zero-temperature, which is assured by taking $m_0^2<0$ (at $D=1+3$) and $\alpha>0$ (at $D=3$). This means, as we have already seen, the nonexistence of a magnetically induced non-thermal transition (at the bulk and with zero chemical potential). 

\begin{figure}
	\centering
	\includegraphics[width=0.7\linewidth]{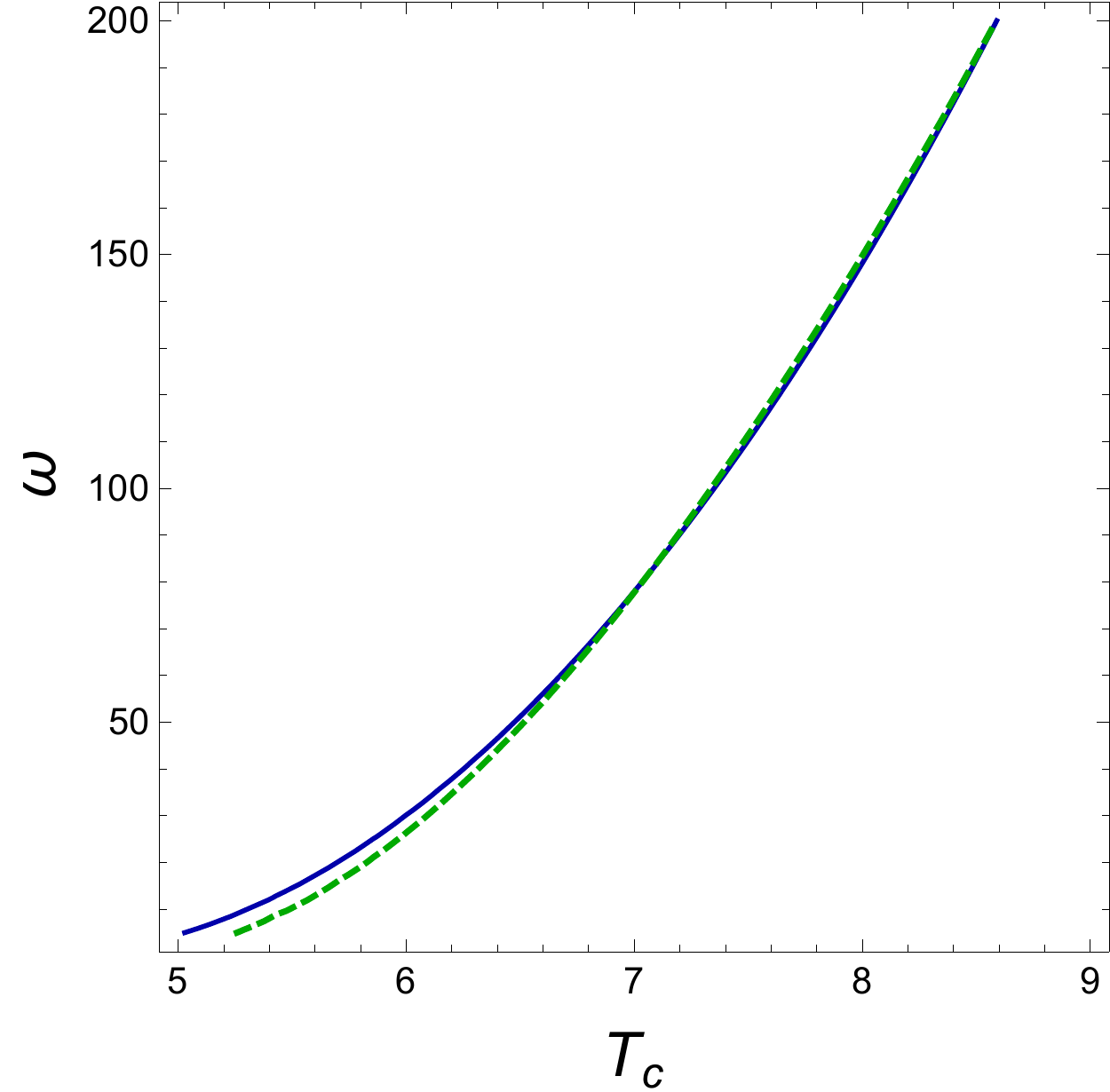}
	\caption{Phase diagram of the critical temperature and external magnetic field at the bulk. Full curve is the $D=3$ model with $T_0 = 4.35$ and $\alpha=0.0569$, the dashed curve is from $D=1+3$ for $m^2=-1$ and series truncated at 50 terms. For both models we used $\lambda=1$.}
	\label{fig:GRAPH_oTvarD}
\end{figure}

To make the comparison clear we start at the bulk ($L\rightarrow\infty$) with zero chemical potential. The phase diagram of the external magnetic field ($\omega$) and critical temperature ($T_c$) is shown in Fig.~\ref{fig:GRAPH_oTvarD} for both $D=3$ (full line) and $D=1+3$ (dashed line) and exhibits a clear increase. This phenomenon is known as magnetic catalysis \cite{Ayala,BoseEinstein01,BoseEinstein02}  and it means that: the increase in $\omega$ favors the system ordering, then it becomes necessary to also increase temperature to restore the disorder/symmetry. Both $D=3$ and $D=1+3$ qualitatively agree; to compute the graph in Fig.~\ref{fig:GRAPH_oTvarD}, we choose some ideal values for $T_0$ and $\alpha$ which allow a quantitative agreement for asymptotic values of the magnetic field with the arbitrary choice of $m_0^2=-1$ and $\lambda=1$. 

\begin{figure}
	\centering
	\includegraphics[width=0.7\linewidth]{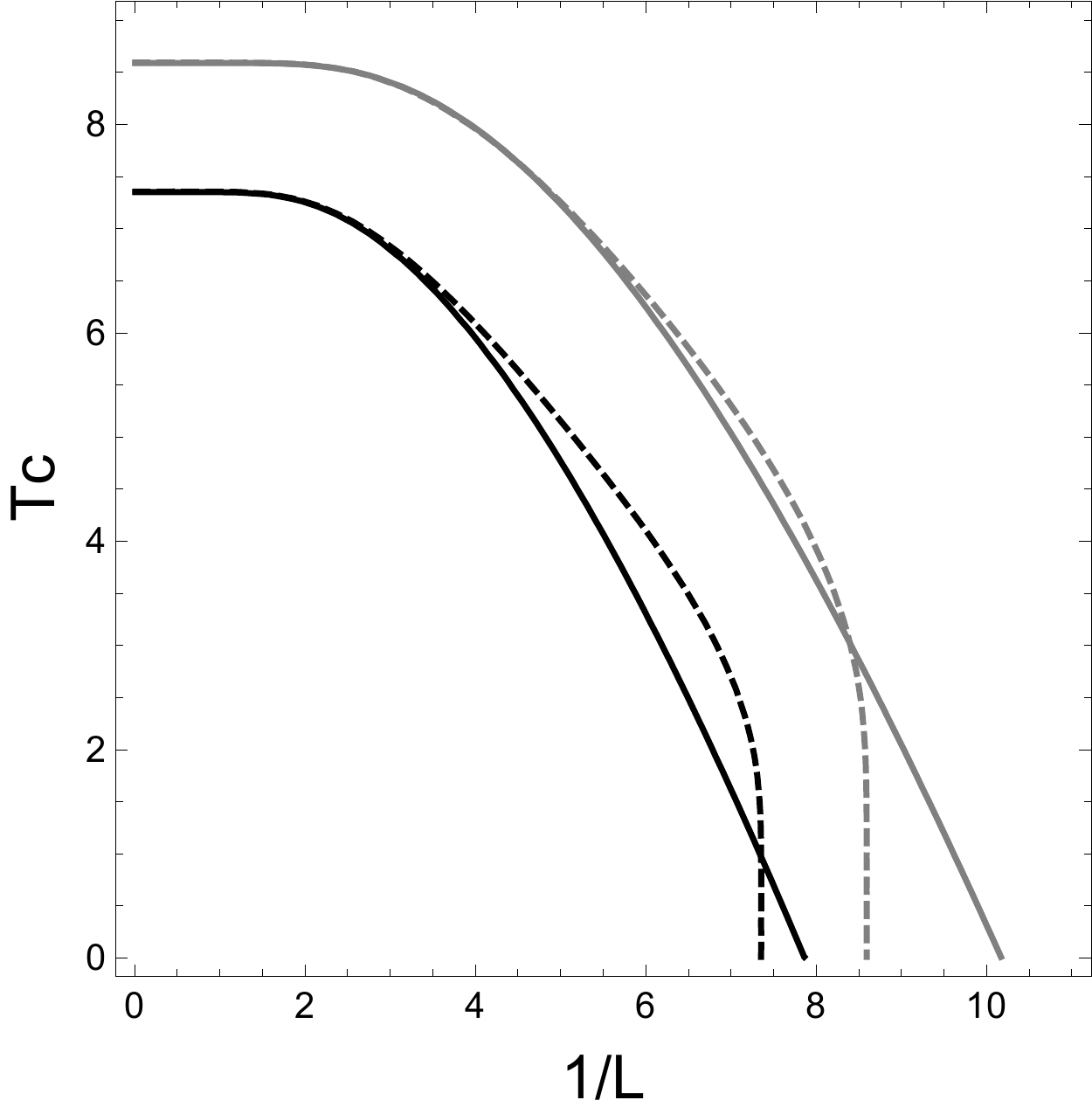}
	\caption{Phase diagram of the critical temperature and inverse system length. Full lines are from $D=3$ (with $T_0=4.35$ and $\alpha=0.0569$) and dashed lines from $D=1+3$ (with $m^2=-1$). For both cases the series where truncated at 10, which was sufficient for strong fields. The black curves have $\omega=100$ and the gray curves $\omega=200$.}
	\label{fig:GRAPH_cTvarD}
\end{figure}

Regarding the arbitrary choice of parameters, it is remarkable that the agreement is sustained not only at the bulk but near it, as shown in Fig.~\ref{fig:GRAPH_cTvarD}. The differences between the models become evident only when the system's size diminishes. This concordance can be seen as a main result, and reveals the existence of a \textit{plateau} near the bulk at which the critical temperature weakly changes. We remark that the ``size" of this \textit{plateau} is increased with the magnetic field.
 
Looking at the graph from another perspective we can see this result as a bound to the validity of the simpler approach of $D=3$: it loses a bit of consistency (quantitatively) as the system width becomes smaller. The exchange symmetry $T \leftrightarrow 1/L$ present at $D=1+3$ is completely lost at $D=3$, and this is reflected in the nonexistence of a \textit{plateau} for a low critical temperature,

However, some features are qualitatively preserved. The\-re is the presence of a minimal length (at which the critical temperature is zero) below which the system is in the disordered phase. This result is in agreement with previous works regarding phase transitions in size-dependent systems~ \cite{PRD2012,Emerson2013,PRD86,Linhares2006,Abreu2013}. Another evident result is that the increase in the external magnetic field not only increases the bulk critical temperature but diminishes the minimal length (both for $D=3$ and $D=1+3$). This means that the increase of the magnetic field allows smaller systems to undergo a phase transition.

\begin{figure}
	\centering
	\includegraphics[width=0.7\linewidth]{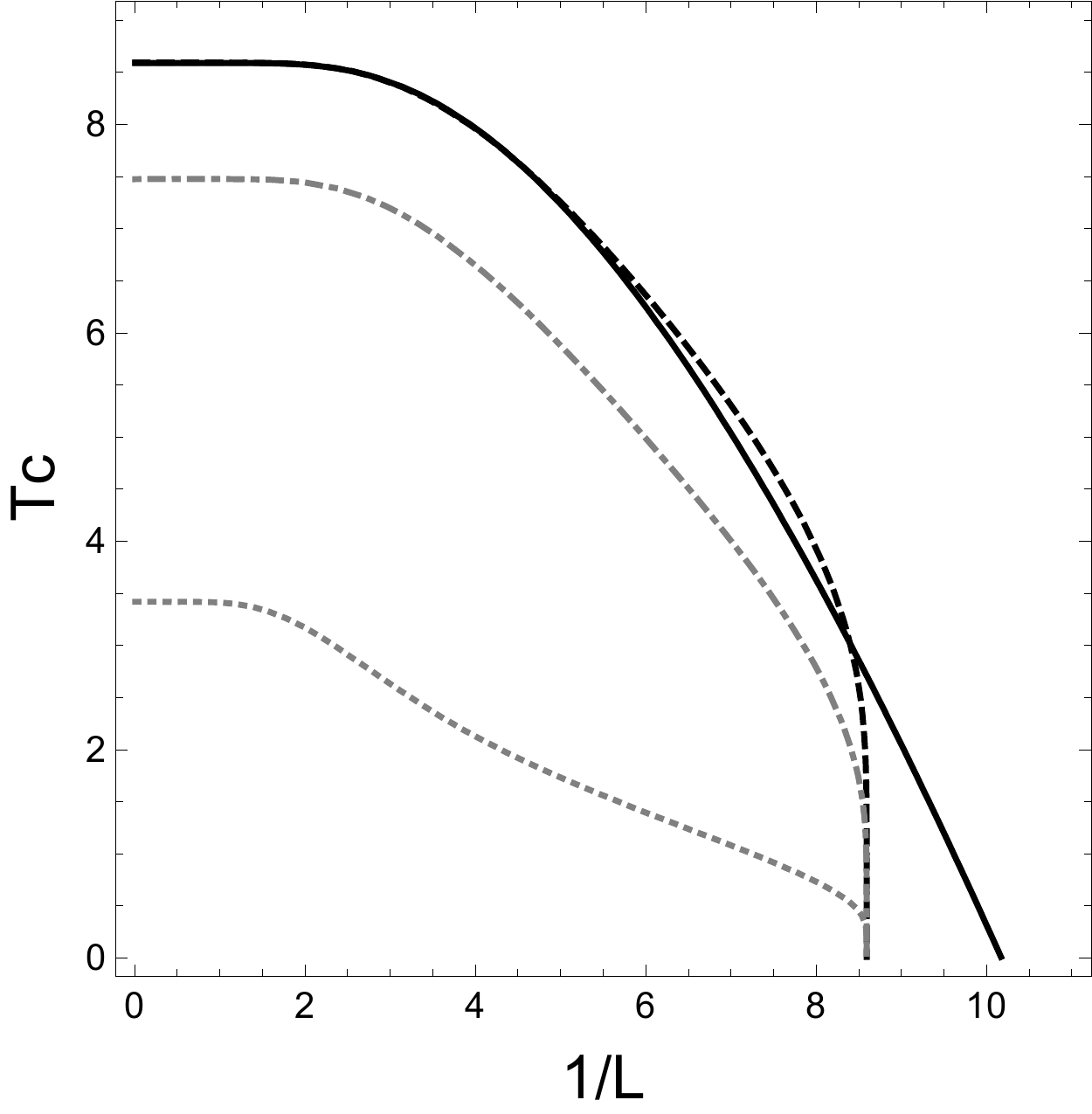}
	\caption{Effect of varying chemical potential $\mu$. Full curve is the $D=3$ model, all others are from $D=1+3$. We have $\mu = 0, 6,12$ respectively the dashed, dot-dashed and dotted curves. The magnetic field is fixed at $\omega=200$. All other parameters are fixed as before. The disordered phase is above the curves}
	\label{fig:GRAPH_cTvarU}
\end{figure}

With the model at $D=1+3$ we can study the chemical potential dependency, as shown at Fig.~\ref{fig:GRAPH_cTvarU}. The remarkable property is that the minimal length is not affected by the chemical potential, while the bulk critical temperature is highly modified. The phase diagram has its shape changed, as the bulk \textit{plateau} diminishes with the increase of the chemical potential. We stress the physical information contained here: the chemical potential does not affect the minimal size of the model but the increase in the chemical potential makes the symmetry restoration easier to achieve.

Another important feature regarding the critical temperature reduction with the chemical potential increase is the existence of a critical point $\mu^\star =\sqrt{m_0^2+\omega}$, where the value of $T_c$ goes to zero regardless of other parameters. Numerically, at this critical point the series diverges and $m_{\text{eff}}^2$ becomes positively divergent in such a way that the phase turns out to be completely disordered. Therefore, the phase diagram is obliterated and only the disordered phase survives. 

\begin{figure}
	\centering
	\includegraphics[width=0.7\linewidth]{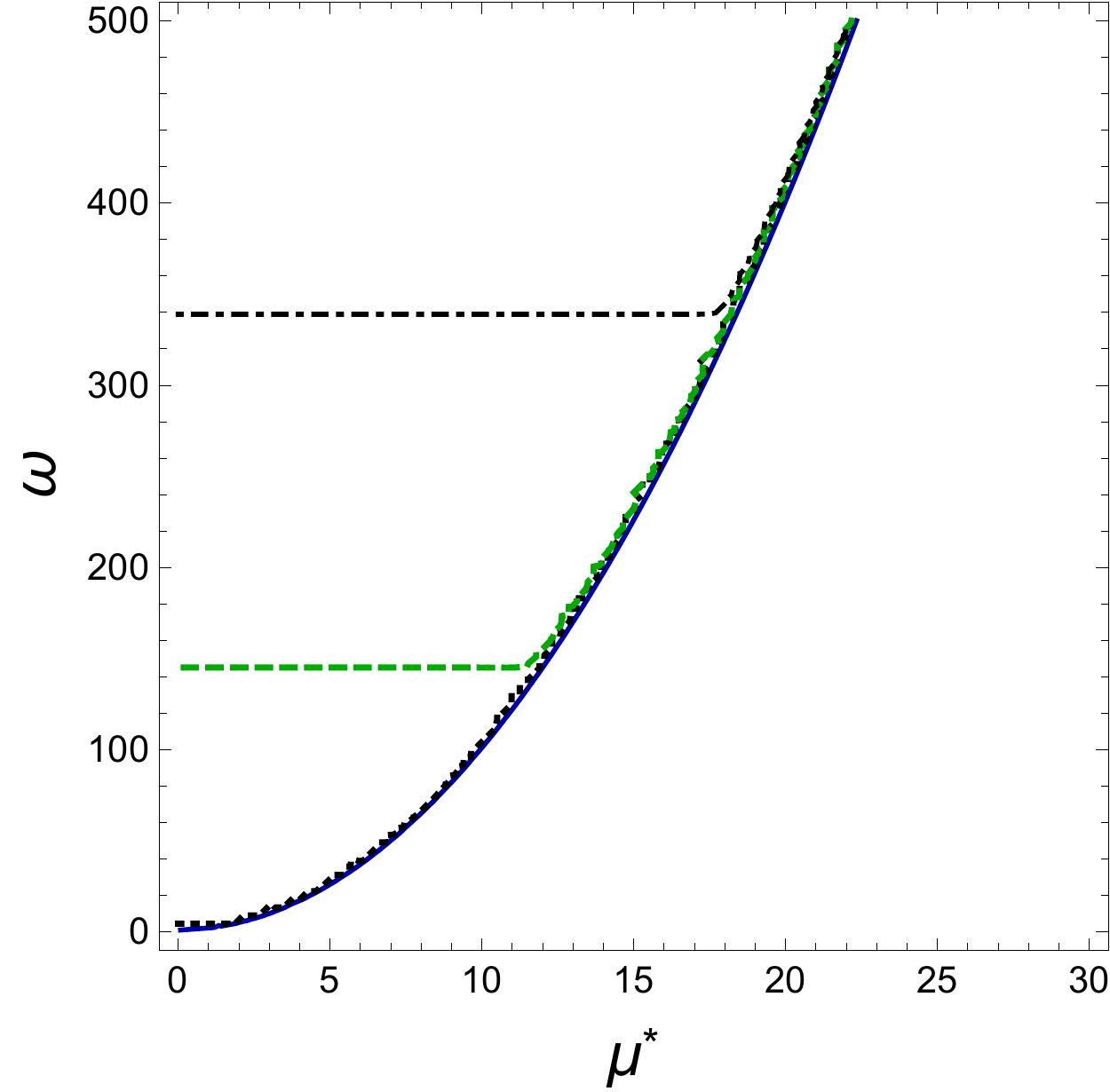}
	\caption{Phase diagram with external field $\omega$ and chemical potential $\mu$ as control parameters. Full line is the bulk $\mu = \sqrt{m^2+\omega}$,  $1/L = 5,8,10$ respectively for the dotted, dashed and dot-dashed curves. Under the curves the phase is disordered, above the curves the phase is ordered and the broken symmetry can be thermally restored. We used $m^2=-1$ and $\lambda=1$.}
	\label{fig:GRAPH_MuOme}
\end{figure}

We must remark that by now we have already introduced the notion of two zero-temperature transitions: the first one is induced by the characteristic length where the critical point is the so-called minimal length $L_{\text{min}}$; while the second one is induced by the chemical potential. A zero-temperature phase diagram depending on $\mu$ and $\omega$ is shown at Fig.~\ref{fig:GRAPH_MuOme} with finite-size effects. The $\omega \times \mu^\star$ behavior near the bulk ($1/L=5$, dotted curve) has a good agreement with the expression $\mu^\star =\sqrt{m^2+\omega}$ (full line) but there exist sensible effects if the system size is below the so-called minimal length $L_{\text{min}}$ as can be seen for $1/L=8,10$ (respectively dashed and dot-dashed curves). This occurs because for $L<L_{\text{min}}$ the system is confined in the symmetric/disordered phase, this introduces the cut seen in Fig.~\ref{fig:GRAPH_MuOme}. Under each curve the phase is disordered, and above it, ordered. We see that we can cross phases using the $\omega,\mu,L$ parameters and still stay at $T_c=0$, so we are undergoing zero-temperature transitions.

\section{Physical discussion}

The monotonically increasing behavior for $T_{c}(\omega)$ for high $\omega$ is associated with the hypothetical phenomenon called magnetic catalysis: strong magnetic fields force the system to stay in the broken/ordered phase, thereby imposing an increasing of the critical temperature. The system considered as the best candidate is a complex Bose gas whose ordered phase is a complex Bose condensate superfluid. This could be related, for example, to a complex pion condensate in a neutron star.

However, a word of caution must be told: the magnetic behavior in this model is a consequence of the particular form in which the thermal and size effects are introduced. Let us take the example of the superconductivity case, also described by using the complex scalar functional of Eq. \eqref{Eq:Action}. In this case magnetism does not favor the superconductivity state, the increase of the external magnetic field decreases the critical temperature and favors the normal state (disordered phase).

In spite of the non-physical magnetic effect predicted by the theory, the predicted finite-size effects could be tested in the superconductivity case (superconducting film). Such ``spurious" magnetic dependence can be absorbed in a redefinition of the mass parameter $m^{2}$ ($D=1+3$ case), allowing to compare with experimental data in a \textit{constant} external magnetic field for a superconducting film. The field model predicts finite-size effects based on our phase transition approach, which can always be tested regardless of the particular magnetic response in the superconducting case.

\section{Conclusions \label{Sec:Conclusion}}
In this article, using the loop expansion at the lowest approximation, we have explored the phase transition phenomena concerning a complex scalar model in the presence of a constant external magnetic field, introducing temperature, finite-size and chemical potential effects. We emphasize that we have not limited ourselves neither to strong magnetic field nor to the high temperature limit. To obtain our results we took into account all Landau levels and also the whole range of temperature ($0\leq \beta^{-1}<\infty$).

We assumed that, considering the bulk (very large system) at zero-temperature and with the magnetic field turned off the system is originally in the ordered/broken phase.

In general, we have observed effects such that the increase in the temperature and chemical potential induces disordering / symmetry restoration, while the increase in the system characteristic length and constant external magnetic field induces ordering / symmetry breaking. The competing effects of these parameters determine the system behavior. 

In the bulk we have observed magnetic catalysis: as the magnetic field grows (inducing more ordering) there is an increase in the critical temperature needed to restore the symmetry. Taking into account size effects, we saw that for tiny systems the critical temperature diminishes and it is possible to define a minimal length below which the phase is disordered and there is no thermal transition. By including the chemical potential, we have seen that, while this does not affect the minimal width, there is a decrease in the critical temperature. We have obtained that at a critical chemical potential the system becomes fully disordered. Based on previous results we have obtained a zero-temperature phase diagram for the model, describing how the magnetic field, chemical potential and finite-size effects can transport the system from the ordered to the disordered phase at zero-temperature. 

We have also shown that the Ginzburg-Landau approximation ($D=3$ model) to introduce temperature directly into the mass term stays qualitatively consistent and, using the models of this article, it only differs from the more suitable finite temperature approach ($D=1+3$ model) when dealing with systems of very small size and low-valued magnetic fields.

\section*{ACKNOWLEDGMENTS}

The authors thank the Brazilian agencies CAPES and CNPq for partial financial support.

\appendix
\section{Zero magnetic field limit\label{Sec:LowOmega}}
In this section we show that the zero magnetic field case is included in our model, although not easily noted. First, to treat the part of Eq.~\eqref{Eq:MassD4} which is independent of the temperature and size we use the asymptotic series of the gamma function 
\begin{equation}
\Gamma(z) \stackrel{z\rightarrow\infty}{\sim} e^{-z} z^{z-1/2}\sqrt{2\pi} \left(1+\frac{1}{12z} + \ldots\right)
\end{equation}
\noindent to obtain, up to the first order correction,
\begin{multline}
\frac{\lambda\omega}{(4\pi)^2} \left\{
-\frac{m^2}{2\omega} \ln\left(\frac{2\pi}{\omega}\right) 
+ \ln\frac{\Gamma\left(\frac{m^2}{2\omega}+\frac{1}{2}\right)}{\sqrt{2\pi}}
\right\}
=\\
\frac{\lambda m^2}{2(4\pi)^2} \left[\ln \frac{m^2}{4\pi} -1\right] -\frac{\lambda  \omega ^2}{12(4\pi) ^2 m^2}
+ \mathcal{O} (\omega^4).
\end{multline}
\noindent We see that the term that survives the limit $\omega\rightarrow0$ is exactly the one that would be obtained if we had considered a model without magnetic field from the beginning.

For the thermal part of Eq.~\eqref{Eq:MassD4},
\begin{equation*}
\frac{\lambda \omega}{(2\pi)^2} \sum_{\ell=0,n=1} K_0\left(\frac{n}{T}\sqrt{m^2+(2\ell+1)\omega}\right),
\end{equation*}
\noindent we use a integral representation of the modified Bessel function of the second kind (Eq. 8.432-6 in \cite{Grad})
\begin{equation}
K_\nu (z) = \frac{1}{2} \left(\frac{z}{2}\right)^\nu \int_0^\infty \frac{e^{-t - \frac{z^2}{4t}}}{t^{\nu+1}} dt \label{Ap:IntRep}
\end{equation}
\noindent so that the sum over the Landau levels is just a geometric series and we obtain
\begin{equation*}
\sum_{\ell=0}^{\infty} \omega K_0\left(\frac{n}{T}\sqrt{m^2+(2\ell+1)\omega}\right)
= \int_0^\infty 
\frac{\omega e^{-t-\frac{m^2 n^2}{4T^2 t}} e^{-\frac{n^2 \omega}{4T^2 t}} }{2t\left(1-e^{-\frac{n^2 \omega}{2T^2 t}}\right)} dt.
\end{equation*}
Expanding for a low-valued external magnetic field we have
\begin{equation*}
\frac{\omega e^{-\frac{n^2 \omega}{4T^2 t}}}{1-e^{-\frac{n^2 \omega}{2T^2 t}}} \approx \frac{2T^2 t}{n^2} - \frac{n^2 \omega^2}{48 T^2 t} + \mathcal{O}(\omega^4),
\end{equation*}
\noindent and we can use again the integral representation of Eq.~\eqref{Ap:IntRep} to obtain
\begin{multline}
\sum_{\ell=0}^{\infty} \omega K_0\left(\frac{n}{T}\sqrt{m^2+(2\ell+1)\omega}\right)
= \frac{m T}{n} K_{-1}\left(\frac{m n}{T}\right) \\- \frac{\omega^2 n m}{24 T} K_1\left(\frac{m n}{T}\right) + \mathcal{O}(\omega^4).
\end{multline}
As $K_{-n}(x) = K_{n}(x)$ we finally see that the zero magnetic field case is included in our model for $D=1+3$.
\begin{multline*}
\frac{\lambda \omega}{(2\pi)^2} \sum_{\ell=0,n=1} K_0\left(\frac{n}{T}\sqrt{m^2+(2\ell+1)\omega}\right) \stackrel{\omega\rightarrow 0}{\approx}\\ \frac{\lambda}{(2\pi)^2} \sum_{n=1} \frac{m T}{n} K_1\left(\frac{n m}{T}\right) + \mathcal{O}(\omega).
\end{multline*}

\FloatBarrier
\bibliography{RefsArticle}

\begin{thebibliography}{38}
\providecommand{\natexlab}[1]{#1}
\providecommand{\url}[1]{\texttt{#1}}
\expandafter\ifx\csname urlstyle\endcsname\relax
  \providecommand{\doi}[1]{doi: #1}\else
  \providecommand{\doi}{doi: \begingroup \urlstyle{rm}\Url}\fi

\bibitem[Cardy(2012)]{cardy}
J.~Cardy.
\newblock \emph{Finite-Size Scaling}.
\newblock Elsevier Science, North Holland, Amsterdam, 2012.

\bibitem[Polchinski(2002)]{polchinski}
J.~Polchinski.
\newblock \emph{Commun. Math. Phys.}, 104:\penalty0 485, 2002.

\bibitem[Panico and Serone(2005)]{panilinha}
G.~Panico and M.~Serone.
\newblock \emph{J. High Energy Phys.}, 05:\penalty0 024, 2005.

\bibitem[Antoniadis(1990)]{pani1}
I.~Antoniadis.
\newblock \emph{Phys. Lett. B}, 246:\penalty0 377, 1990.

\bibitem[Burdman and Nomura(2003)]{pani3}
G.~Burdman and Y.~Nomura.
\newblock \emph{Nucl. Phys. B}, 656:\penalty0 3, 2003.

\bibitem[Arkani-Hamed et~al.(2002{\natexlab{a}})Arkani-Hamed, Cohen, Gregoire,
  Katz, Nelson, and Walker]{pani6}
N.~Arkani-Hamed, A.~G. Cohen, T.~Gregoire, E.~Katz, A.~E. Nelson, and J.G.
  Walker.
\newblock \emph{J. High Energy Phys.}, 08:\penalty0 021, 2002{\natexlab{a}}.

\bibitem[Arkani-Hamed et~al.(2002{\natexlab{b}})Arkani-Hamed, Cohen, Katz, and
  Nelson]{pani6b}
N.~Arkani-Hamed, A.~G. Cohen, E.~Katz, and A.~E. Nelson.
\newblock \emph{J. High Energy Phys.}, 07:\penalty0 034, 2002{\natexlab{b}}.

\bibitem[Arkani-Hamed et~al.(2001)Arkani-Hamed, Cohen, and Georgi]{pani7}
N.~Arkani-Hamed, A.~G. Cohen, and H.~Georgi.
\newblock \emph{Phys. Lett. B}, 513:\penalty0 232, 2001.

\bibitem[Roy and Bander(2009)]{(g-2)NPB}
A.~J. Roy and M.~Bander.
\newblock \emph{Nucl. Phys. B}, 811:\penalty0 353, 2009.

\bibitem[Ttira et~al.(2010)Ttira, Fosco, Malbouisson, and Roditi]{claudio}
C.~Ccapa Ttira, C.~D. Fosco, A.~P.~C. Malbouisson, and I.~Roditi.
\newblock \emph{Phys. Rev. A}, 81:\penalty0 032116, 2010.

\bibitem[Khanna et~al.(2014)Khanna, Malbouisson, Malbouisson, and
  Santana]{PhysReport}
F.~C. Khanna, A.~P.~C. Malbouisson, J.~M.~C. Malbouisson, and A.~E. Santana.
\newblock \emph{Phys. Rep.}, 539:\penalty0 135--224, 2014.

\bibitem[Khanna et~al.(2011)Khanna, Malbouisson, Malbouisson, and
  Santana]{AOP11}
F.~C. Khanna, A.~P.~C. Malbouisson, J.~M.~C. Malbouisson, and A.~E. Santana.
\newblock \emph{Ann. Phys.}, 326:\penalty0 2364, 2011.

\bibitem[Khanna et~al.(2009{\natexlab{a}})Khanna, Malbouisson, Malbouisson, and
  Santana]{AOP09}
F.~C. Khanna, A.~P.~C. Malbouisson, J.~M.~C. Malbouisson, and A.~E. Santana.
\newblock \emph{Ann. Phys.}, 324:\penalty0 1931, 2009{\natexlab{a}}.

\bibitem[Khanna et~al.(2009{\natexlab{b}})Khanna, Malbouisson, Malbouisson, and
  Santana]{TheBook}
F.~C. Khanna, A.~P.~C. Malbouisson, J.~M.~C. Malbouisson, and A.~E. Santana.
\newblock \emph{Thermal Quantum Field Theory: Algebraic Aspects and
  Applications}.
\newblock World Scientific, Singapore, 2009{\natexlab{b}}.

\bibitem[Malbouisson et~al.(2002)Malbouisson, Malbouisson, and
  Santana]{NuclPhys2002}
A.~P.~C. Malbouisson, J.~M.~C. Malbouisson, and A.~E. Santana.
\newblock \emph{Nucl. Phys. B}, 631:\penalty0 83, 2002.

\bibitem[Abreu et~al.(2006)Abreu, Gomes, and da~Silva]{Abreu2006}
L.~M. Abreu, M.~Gomes, and A~J. da~Silva.
\newblock \emph{Phys. Lett. B}, 642:\penalty0 551, 2006.

\bibitem[Abreu et~al.(2009)Abreu, Linhares, Malbouisson, Malbouisson, and
  Santana]{Abreu2009}
L.~M. Abreu, C.~A. Linhares, A.~P.~C. Malbouisson, J.~M.~C. Malbouisson, and
  A.~E. Santana.
\newblock \emph{Nucl. Phys. B}, 819:\penalty0 127, 2009.

\bibitem[Linhares et~al.(2012{\natexlab{a}})Linhares, Malbouisson, and
  Roditi]{EPL2012}
C.~A. Linhares, A.~P.~C. Malbouisson, and I.~Roditi.
\newblock \emph{EPL}, 98:\penalty0 41001, 2012{\natexlab{a}}.

\bibitem[Khanna et~al.(2012)Khanna, Malbouisson, Malbouisson, and
  Santana]{PRD2012}
F.~C. Khanna, A.~P.~C. Malbouisson, J.~M.~C. Malbouisson, and A.~E. Santana.
\newblock \emph{Phys. Rev. D.}, 85:\penalty0 085015, 2012.

\bibitem[Corr\^ea et~al.(2013)Corr\^ea, Linhares, and Malbouisson]{Emerson2013}
E.~B.~S. Corr\^ea, C.~A. Linhares, and A.P.C Malbouisson.
\newblock \emph{Phys. Lett. A}, 377:\penalty0 1984, 2013.

\bibitem[Linhares et~al.(2012{\natexlab{b}})Linhares, Malbouisson, Malbouisson,
  and Roditi]{PRD86}
C.~A. Linhares, A.~P.~C. Malbouisson, J.~M.~C. Malbouisson, and I.~Roditi.
\newblock \emph{Phys. Rev. D}, 86:\penalty0 105022, 2012{\natexlab{b}}.

\bibitem[Linhares et~al.(2006)Linhares, Malbouisson, Milla, and
  Roditi]{Linhares2006}
C.~A. Linhares, A.~P.~C. Malbouisson, Y.~W. Milla, and I.~Roditi.
\newblock \emph{Phys. Rev. B}, 73:\penalty0 214525, 2006.

\bibitem[Abreu et~al.(2013)Abreu, Linhares, Malbouisson, and
  Malbouisson]{Abreu2013}
L.~M. Abreu, C.~A. Linhares, A.~P.~C. Malbouisson, and J.~M.~C. Malbouisson.
\newblock \emph{Phys. Rev. D}, 88:\penalty0 107701, 2013.

\bibitem[Ayala et~al.(2012)Ayala, Loewe, Rojas, and Villavicencio]{Ayala}
A.~Ayala, M.~Loewe, J.~C. Rojas, and C.~Villavicencio.
\newblock \emph{Phys. Rev. D}, 86:\penalty0 076006, 2012.

\bibitem[Feng et~al.(2016)Feng, Hou, Ren, and Wu]{BoseEinstein01}
B.~Feng, D.~Hou, H.~Ren, and P.~Wu.
\newblock \emph{Phys. Rev. D}, 93:\penalty0 085019, 2016.

\bibitem[Loewe et~al.(2014)Loewe, Villavicencio, and Zamora]{BoseEinstein02}
M.~Loewe, C.~Villavicencio, and R.~Zamora.
\newblock \emph{Phys. Rev. D}, 89:\penalty0 016004, 2014.

\bibitem[Malbouisson(2002)]{PRB2002}
A.~P.~C. Malbouisson.
\newblock \emph{Phys. Rev. B}, 66:\penalty0 092502, 2002.

\bibitem[Abreu et~al.(2004)Abreu, de{ }Calan, and Malbouisson]{Calan2004}
L.~M. Abreu, C.~de{ }Calan, and A.~P.~C. Malbouisson.
\newblock \emph{Phys. Lett. A}, 333:\penalty0 316, 2004.

\bibitem[Abreu et~al.(2003)Abreu, Malbouisson, Malbouisson, and
  Santana]{Abreu2003}
L.~M. Abreu, A.~P.~C. Malbouisson, J.~M.~C. Malbouisson, and A.~E. Santana.
\newblock \emph{Phys. Rev. B}, 67:\penalty0 21250, 2003.

\bibitem[Calza et~al.(2016)Calza, Cardoso, Cardoso, and Linhares]{Calza2016}
T.~C.~A. Calza, F.~L. Cardoso, L.~G. Cardoso, and C.~A. Linhares.
\newblock \emph{Mod. Phys. Lett. A}, 31:\penalty0 1650227, 2016.

\bibitem[Lawrie(1997)]{Lawrie}
I.~D. Lawrie.
\newblock \emph{Phys. Rev. Lett.}, 79:\penalty0 131, 1997.

\bibitem[Bollini and Giambiagi(1972)]{DimReg}
C.~G. Bollini and J.~J. Giambiagi.
\newblock \emph{Il Nuovo Cimento}, 12:\penalty0 20, 1972.

\bibitem['t{ }Hooft and Veltman(1972)]{DimReg2}
G.~'t{ }Hooft and M.~Veltman.
\newblock \emph{Nucl. Phys. B}, 44:\penalty0 189, 1972.

\bibitem[Elizalde and Romeo(1989)]{ElizaldeRomeo}
E~Elizalde and A~Romeo.
\newblock \emph{J. Math. Phys.}, 30:\penalty0 1133, 1989.

\bibitem[Elizalde(2012)]{elizaldebook}
E.~Elizalde.
\newblock \emph{Ten Physical Applications of Spectral Zeta Functions}.
\newblock Springer, Berlin Heidelberg, 2012.

\bibitem[Continentino and Ferreira(2007)]{Mucio2007}
M.~A. Continentino and A.~S. Ferreira.
\newblock \emph{J. Magn. Magn. Mater.}, 310:\penalty0 828, 2007.

\bibitem[Continentino(2001)]{Muciobook}
M.~A. Continentino.
\newblock \emph{Quantum Scaling in Many-Body Systems}.
\newblock World Scientific, Singapore, 2001.

\bibitem[Gradshteyn and Ryzhik(2014)]{Grad}
I.~S. Gradshteyn and I.~M. Ryzhik.
\newblock \emph{Table of Integrals, Series, and Products}.
\newblock Elsevier Academic Press, 7th edition, 2014.

\end{thebibliography}

\end{document}